\documentclass[
  aip,           
  pop,           
  amsmath,amssymb,twocolumn,
  reprint,
]{revtex4-1}
\makeatletter
\providecommand{\bibinfo}[2]{#2}%
\providecommand{\bibfnamefont}[1]{#1}%
\providecommand{\bibnamefont}[1]{#1}%
\providecommand{\eprint}[2][]{\url{#2}}%
\providecommand{\Eprint}[2][]{\url{#2}}%
  \providecommand{\aap}{Astron.\ Astrophys.}
\makeatother

\usepackage{graphicx}
\usepackage{dcolumn}
\usepackage{bm}
\usepackage{letterspace}
\usepackage[utf8]{inputenc}
\usepackage[T1]{fontenc}
\usepackage{mathptmx}
\usepackage{amsmath,amssymb}
\usepackage{etoolbox}
\usepackage{comment}
\usepackage{setspace}
\usepackage{caption}
\usepackage{xcolor}

\usepackage[normalem]{ulem}

\makeatletter
\def\@email#1#2{%
 \endgroup
 \patchcmd{\titleblock@produce}
  {\frontmatter@RRAPformat}
  {\frontmatter@RRAPformat{\produce@RRAP{*#1\href{mailto:#2}{#2}}}\frontmatter@RRAPformat}
  {}{}
}%
\makeatother
\usepackage[colorlinks=true,
            linkcolor=blue,
            citecolor=blue,
            urlcolor=blue]{hyperref}
\begin{document}

\preprint{Phys.\ Plasmas 32, 042111 (2025)}

\title[Dispersion properties of {\textcolor{black}{neutron star}} magnetospheric plasmas with relativistic kappa distribution]{Dispersion properties of {\textcolor{black}{neutron star}} magnetospheric plasmas with relativistic kappa distribution}
\author{M. Mousavi}
\email{mohaddeseh.mousavi@stud.uni-due.de.}

\affiliation{ Faculty of Physics, University of Duisburg-Essen, Lotharstraße 1, 47057 Duisburg, Germany}

\author{J. Benáček}
\email{jan.benacek@uni-potsdam.de.}
\affiliation{ 
Institute for Physics and Astronomy, University of Potsdam, 14476 Potsdam, Germany 
}%
\affiliation{Center for Astronomy and Astrophysics, Technical University of Berlin, 10623 Berlin, Germany}


\begin{abstract}
The \textcolor{black}{various distribution functions} can encompass the diverse characteristics of the magnetospheric plasma of surrounding neutron stars in both hot and cold environments; however, the Maxwell-Jüttner distribution is so far widely used to characterize these plasmas.
We aim to analyze the linear dispersion properties yielded from the relativistic kinetic dispersion relation for the {\textcolor{black}{neutron star}} magnetospheric plasmas. 
We developed a numerical dispersion solver to investigate plasmas with arbitrary velocity distributions \textcolor{black}{and focus on the comparison of} relativistic kappa and Maxwell-Jüttner distribution functions \textcolor{black}{as analytical representatives}.
By considering different kappa distribution indices and using analytical and numerical approaches, the dispersion properties of the kappa and Maxwell-Jüttner distributions \textcolor{black}{approach each other for} high wave numbers and low temperatures, indicating that the choice of distribution functions has little effect on \textcolor{black}{high} wave numbers $ck/\omega_p \gg 1$ and high inverse temperatures $\rho=100$.
However, each distribution function exhibits unique \textcolor{black}{yet} complementary properties in semi-relativistic to relativistic inverse temperatures $\rho \leq 10^{-1}$ and at lower wave numbers $ck/\omega_p\leq 1$.  This highlights the necessity of utilizing such a dispersion solver for these \textcolor{black}{wave numbers} to properly comprehend the dispersion properties of the neutron star magnetospheric plasmas.
\end{abstract}

\maketitle

\section{\label{sec:intro}INTRODUCTION}

Understanding the dispersion properties of neutron star magnetospheric plasmas holds the key to unraveling the wide range of electromagnetic radiation emitted by these celestial objects, ranging from radio waves to X-rays and gamma rays~\citep{grenier2015gamma}. The mechanisms behind neutron star emissions remains partially elusive, with indications pointing toward the crucial involvement of plasma on kinetic scales within their magnetospheres~\citep{philippov2022pulsar}. 

Exploring the dispersion properties of neutron star magnetospheric plasmas, particularly the relativistic kappa velocity distribution in neutron star magnetospheres, is a crucial prerequisite for unraveling the effects of electromagnetic wave propagation and the growth of instabilities within these environments. Understanding the dispersion properties offers invaluable insight into the origin, propagation, and processable mechanisms of electromagnetic waves and particles at kinetic scales~\citep{kazbegi1987radio,melrose1989book}. Developing a generally usable theoretical framework may not only aid in deciphering observational data, but also deepen our understanding of the underlying physics governing neutron star magnetospheric plasmas~\citep{manthei2021refining}.

In neutron star magnetospheres, the plasma can be modeled using plasma waves characterized by the frequency ($\omega$) and wave number ($k$). For wave solutions, the light line $\omega = ck$ divides the plane into areas of ``superluminal'' and ``subluminal'' propagation~\citep{bortman2001phase}.

The dispersion properties of neutron star magnetospheric plasmas for different distribution functions have been extensively studied in recent years~\citep{melrose2017coherent,fasano2019constraining,2019JPlPh..85c9005R}. The low plasma density in these environments results in a collisionless plasma, where particle interactions are governed by collective effects rather than direct Coulomb collisions. As a result, the plasma can exhibit both thermal equilibrium and nonthermal characteristics, depending on the dominant physical processes~\citep{baumjohann2012basic,lazar2021kappa}. In addition, the ions present in the plasma are negligible because they are mostly replaced by positrons in pair-dominated plasmas and due to their significantly higher mass. The Maxwell-Jüttner distribution, while not fully representative of the pair-cascade plasma in pulsars, has been employed as an analytical model due to its simplicity and mathematical tractability to explore dispersive properties~\citep{2019JPlPh..85c9005R}. Simulations, although insightful, often do not fully resolve the velocity space at all relevant scales, making simpler distribution profiles reasonable approximations in certain regions of the neutron star magnetosphere~\citep{Benacek2024b}. In particular, on closed field lines where plasma dynamics is more stable compared to the complex polar cap regions~\citep{arendt2002pair}, the Maxwell-Jüttner distribution can serve as a suitable first-order approximation. By expanding our focus to neutron star magnetospheres, our work provides insights applicable to a wider range of astrophysical environments, where distribution functions such as Maxwell-Jüttner or kappa distributions may be relevant.

Furthermore, observations have revealed that neutron star magnetospheric plasmas possess complex dispersion properties due to their nonthermal nature. For example, \citet{briozzo2023analytical} worked on the propagation of light rays resulting in the chromatic analysis, and \citet{akbari2021langmuir} included the electrostatic dispersion relation, Langmuir wave excitation, and beam distribution stability. \citet{mikhaylenko2021thermal} studied the thermal effects on the dispersion properties of four electromagnetic wave modes propagating in the pulsar magnetosphere. The kappa velocity distribution has been suggested as a more credible model to describe neutron star magnetospheric plasmas by \citet{livadiotis2022physical}.

The presence of power-law tails in X-ray spectra, often associated with synchrotron radiation from high-energy electrons, is observed across various astrophysical environments, including intrabinary shocks in spider pulsars, ultraluminous X-ray pulsars, and pulsar wind nebulae, where spectral modeling reveals hard photon indices and high-energy cutoffs indicative of nonthermal particle acceleration processes~\citep{kargaltsev2024hard, kumar2025spectral, cortes2025particle}.

The kappa distribution has high-energy tails in the distribution function profile, which deviates from the thermal Maxwell-Boltzmann or Maxwell-Jüttner distributions so far and is commonly observed in space and astrophysical plasmas~\citep{yoon2018weak,shapiro1988astrophysical}. This distribution bridges the gap between Maxwellian and power-law behaviors, making it particularly valuable for studying non-thermal features in highly magnetized astrophysical plasmas. It is especially suited for modeling stable plasmas outside the polar caps of neutron star magnetospheres, where distribution functions are likely broader peaked and exhibit power-law tails. One of the significant findings in the dispersion properties of neutron star magnetospheric plasmas with relativistic velocity distributions is that the thermal properties of the kappa distribution lead to various types of waves, including Langmuir, electromagnetic, and Alfvén waves~\citep{melrose2017coherent}.

The dispersion properties of these waves depend on the kappa parameter, which characterizes the presence of high-velocity particles in the plasma distribution. The kappa parameter significantly influences wave properties and their possible impact on the background plasma~\citep{benavcek2019growth,10.1093/mnras/staa2280}.

This study delves into the relativistic nature of the kappa distribution and examines how it influences wave dispersion and the behavior of relativistically hot plasma within the neutron star magnetosphere. Specifically, we investigated a comparison of the kappa dispersion properties with those of the Maxwell-Jüttner distributions, both viewed as comprehensive concepts. We take advantage of the distribution function prescriptions for modeling dispersion properties in relativistic magnetospheric neutron star plasmas for thermal and nonthermal particle populations, proposed in \citet{2019JPlPh..85c9005R} and \citet{livadiotis2013understanding}. Some of the existing simulations focus on the polar cap regions of neutron star magnetospheres, where pair discharges dominate and distribution functions exhibit high-energy population with complex features, the plasmas confined to the closed field lines of the magnetosphere likely show more thermal behavior and are less dominated by extreme energy populations. Our study does not constrain itself to polar cap regions, but instead aims to examine these more stable plasmas. The kappa distribution is particularly suitable for modeling such environments because of its ability to describe power-law tails of the particle energy distributions in the magnetospheric plasma.
By choosing these two distributions, our aim is to provide both a comparative baseline and an example of nonthermal effects, highlighting the flexibility and utility of our approach. Each distribution influences the characteristics of astrophysical plasma systems in a unique way. By employing numerical and analytical methods, we aim to extend the dispersion model and scrutinize various parameters of the kappa distribution to uncover the complementarity while highlighting differences in comparison to the Maxwell-Jüttner distribution.

The paper is structured as follows. In the Methods section, we encompass analytical calculations of the dispersion properties for the distribution functions, and describe the developed numerical model. Subsequently, the Results section presents the findings obtained through these methods, elucidating the comparative analysis between the kappa and Maxwell-Jüttner distributions. Following this, the Discussion section delves into the interpretation and implications of the results, highlighting the significance of the shown differences and complementarities. Finally, the paper concludes with a brief yet comprehensive summary in the Conclusion section, which summarizes the key takeaways and avenues for further research in this field.

\section{METHODS}
\subsection{Analytical model}
\begin{figure}
\raggedleft
\includegraphics[width=\columnwidth]{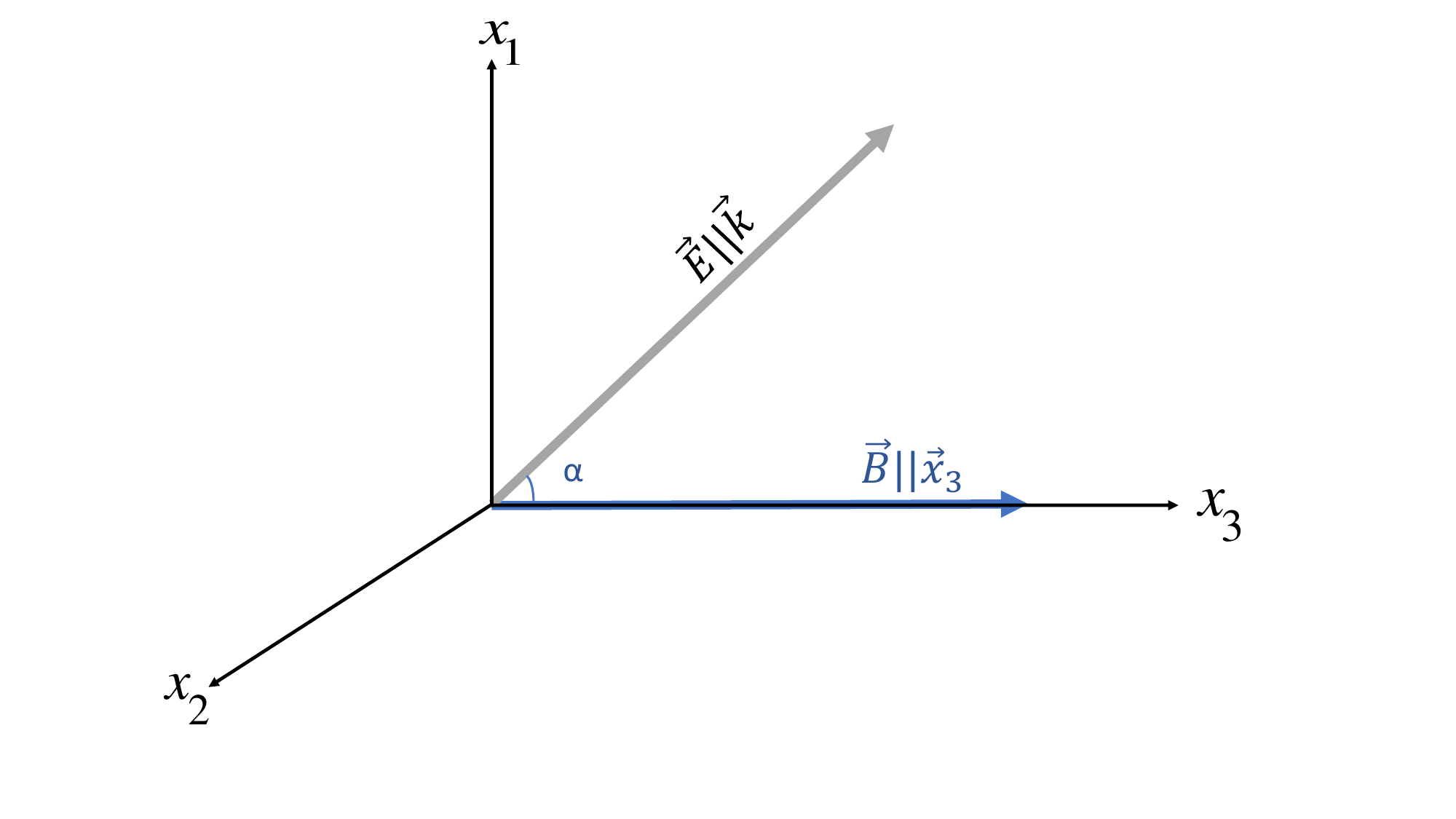}
	\caption{showing of the wave vector ($\boldsymbol{k}$) propagation, magnetic field ($B$) orientation, and electric field ($E$) configuration, with the propagation angle ($\alpha$) relative to the magnetic field.}
	\label{fig:drawing}
\end{figure}

To study the wave properties and particle behavior, we assume that the plasma has a uniform and constant temperature throughout the background and the dispersion relation can be obtained by calculating the 1D dielectric tensor. We utilize the theory of weakly damped waves proposed by \citet{melrose1999relativistic} and \citet{rafat2019wave}. {\textcolor{black}{As Fig.\ref{fig:drawing} shows the geometric setup, the magnetic field vector ($\boldsymbol{B}$), wave vector ($\boldsymbol{k}$), and electric field vector ($\boldsymbol{E}$), and the propagation angle ($\alpha$) is denoted in a uniform medium.}} Formulating based on \citet{melrose2008quantum} approach for a strongly magnetized plasma modeled for a certain distribution, we utilize the wave equation $\boldsymbol{\Lambda}_{ij}(\omega, \mathbf{k}) \mathbf{e}_j=0$, where $\boldsymbol{\Lambda}_{ij}$ is the plasma dispersion tensor and $\boldsymbol{e}_j$ is the electric field amplitude. To solve for the wave equation, the determinant of $\boldsymbol{\Lambda}_{ij}$ is calculated, which can be expressed as
\begin{equation}\label{eq:1}
    \mathrm{det}\boldsymbol{\Lambda}_{ij} = \boldsymbol{\Lambda}_{22}(\boldsymbol{\Lambda}_{11}\boldsymbol{\Lambda}_{33}-\boldsymbol{\Lambda}_{13}^2).
\end{equation}

The electromagnetic wave solutions are then expressed as follows {\textcolor{black}{
\begin{equation}
\boldsymbol{\Lambda}_{11} \times \boldsymbol{\Lambda}_{33} - \boldsymbol{\Lambda}_{13}^2 = 0, \qquad
\boldsymbol{\Lambda}_{22} = 0
\end{equation}
}}where $\boldsymbol{\Lambda}_{33}$ represents the modes parallel to the magnetic field direction.

The plasma response tensor is given by the equations below. The first line represents a dispersion relation for wave propagation, and the second line defines the explicit form of the response tensor showing more contributions to the electromagnetic field, where {$c$} is the speed of light.
\begin{align}\label{eq:2}
	&\boldsymbol{\Lambda}_{ij}(\omega, \boldsymbol{k}) \boldsymbol{e}_j =0,\\\notag
	&\boldsymbol{\Lambda}_{ij}(\omega, \boldsymbol{k}) = \frac{c^2(\boldsymbol{k}_i \boldsymbol{k}_j -\lvert \boldsymbol{k}^2\rvert\delta_{ij})}{\omega^2}+\boldsymbol{k}_{ij}(\omega, \boldsymbol{k}).
\end{align}
While $\boldsymbol{k}_i$ and $\boldsymbol{k}_j$ are components of the wave vector, and $\boldsymbol{k}_{ij}$ tensor components depend on frequency and wave number. To define the theory in terms of the relativistic plasma dispersion functions
(RPDF) in the rest of the paper, we solve the wave equation tensor $\boldsymbol{\Lambda}_{ij}$ for the $\boldsymbol{k}_{ij}$ components which lead to the normalized dispersion equation denoted as $W({z})$  (see \citet{melrose1986instabilities}, Eq (2.30) \textcolor{black}{we use a different notation}) , where ${z}={\omega}/{{k}_{||}c}$ is the wave phase \textcolor{black}{velocity}, ${k}_{||}$ is the wave number along the magnetic field, and $\omega$ is the wave frequency. 
By assuming a one-dimensional group velocity along the 3-axis, the modified version of the dispersion relation integral $W({z})$ is a function defined as \citep{fried1961sd,melrose1999relativistic}
\begin{equation}\label{eq:3}
	W({z}) = \tfrac{1}{n}\int d{u} \frac{1}{{\beta} - {z}}\frac{dg({u})}{d{u}},
\end{equation}
where $n=\int_{-\infty}^{+\infty} d{u} \,g({u}) = 1$ and $g({u})$ is the 1D particle distribution function and ${\beta}$ is particle velocity in units of light speed $c$.

The denominator of Eq.(\ref{eq:3}) {\textcolor{black}{produces}} the Cerenkov resonance condition for ${\beta}={z}$, leading to a singularity in the calculation of the integral. The velocity, denoted as ${u}$ is expressed in terms of the Lorentz factor ${\gamma}$ using ${u} = {\gamma}{\beta}$. Here, ${\gamma}$ is defined as ${\gamma}=1\diagup (1-{\beta}^2)^{1/2}$, and ${\beta} = {v}/c$ represents the particle speed, expressed in units of the speed of light $c$.
The function $W({z})$, can be computed for ${z}\leq1$ as

\begin{equation}\label{eq:4}
	W({z}) = \lim_{\delta \rightarrow 0 } \left( \int_{-1} ^ {{z} - \delta} + \int_{{z} - \delta} ^ {{z}+ \delta} + \int_{{z} + \delta}^{1} \right) d{\beta} \frac{1}{{\beta} -{z} + \epsilon} \frac{dg({u})}{d{\beta}}.
\end{equation}
Where $\epsilon \lll 1$ is a parameter for assuring numerical integrability. The Eq. (\ref{eq:4}) leads to the Cauchy principal value. The integral is taken over three intervals, each {\textcolor{black}{having}} a small interval of width $\delta$ centered at $z$. The value of $W({z})$ is the limit of this integral as $\delta$ approaches zero.

As long as the distribution function does not grow rapidly, $W({z})$ is singular at ${\beta} = {z} $ for $\lvert {z} \lvert <1$. The Laplace transform can be used to find a point ${z}$

\begin{equation}\label{eq:5}
	{z} = \frac{\omega}{{k}_{\parallel}c} \hspace{0.3 cm} implies \hspace{0.3 cm} \left\{
	\begin{tabular}{@{}c@{}}
		${z} \rightarrow {z}+i0 \hspace{0.3 cm} for \hspace{0.3 cm} {k}_{\parallel}>0$,\\
		${z} \rightarrow {z}-i0 \hspace{0.3 cm} for \hspace{0.3 cm} {k}_{\parallel}<0$.
	\end{tabular}
	\right.
\end{equation}

\subsubsection{Kappa distribution}\label{sec:kappa}

{\textcolor{black}{The kappa distribution function can be used to model the distribution of particles \citep{roussos2022kappa}.}} {\textcolor{black}{kappa distribution is basically considered to explore how power-law tails, which are common in astrophysical plasmas, influence dispersion properties.} 
Based on the notation in \citet{livadiotis2013understanding}, we have 
\textcolor{black}{
\begin{align*}
	&g({u}; \theta; \kappa_0) =
	\left( \pi \kappa_0 \theta^2 \right)^{-\tfrac{1}{2}} \frac{\Gamma(\kappa_0 +1+ \frac{1}{2})}{\Gamma(\kappa_0 - \tfrac{1}{2})} \left( 1+ \frac{1}{\kappa_0} \frac{{u}^2}{\theta^2} \right)^{-\kappa_0 -\tfrac{3}{2}},
\end{align*}}
where the distribution function is denoted by \textcolor{black}{$g({u}; \theta; \kappa_0)$}, ${u}$ is the velocity of the particles, $\theta$ is the temperature parameter \textcolor{black}{which is normalized to $mc^2$}. \textcolor{black}{In fact, the temperature parameter $\theta$ is related to momentum, which is Lorentz invariant, rather than velocity. While the kappa distribution itself is not Lorentz invariant, our study is conducted in the bulk reference frame, where $\theta$ is defined. This simplifies the analysis and enables characterization of the distribution within a well-defined reference frame.} $\kappa_0$ is the invariant kappa index parameter, with the value $\kappa = \kappa_0 +{3}/{2}$ , and $\Gamma(x)$ is the gamma function. \textcolor{black}{The slope of the power law tail in the kappa distribution is relevant to $-\kappa$, with lower $\kappa$ values leading to extended distribution tails. This study focuses on analyzing the properties of these tails, particularly the transition from Maxwell-Jüttner to kappa distributions, to highlight the influence of nonthermal features on dispersion properties.}

The kappa distribution function is defined by two main parameters: the index $\kappa$ and the thermal velocity $\theta$. \textcolor{black}{The kappa parameter}, which has values $>\frac{3}{2}$. When $\kappa = \infty$, the plasma is in thermal equilibrium. In contrast, when $\kappa$ approaches $3/2$, the plasma is in a state of ``anti-equilibrium,`` which is a unique state characterized by a wide range of properties.}  Both the equilibrium and anti-equilibrium states exhibit \textcolor{black}{consistent} behavior that is independent of system dimensionality. At equilibrium, the kappa distribution reverts to the Maxwell-Boltzmann velocity distribution. {\textcolor{black}{The anti-equilibrium state refers to a specific regime of the kappa distribution where it approaches a power-law distribution. However, it is important to note that power-law behavior can also emerge for arbitrary and finite values of kappa, not exclusively in the anti-equilibrium limit }}\citep{livadiotis2013understanding,fisk2006common}.

The distribution is normalized in such a way that the integral over all velocities is equal to unity.
When integrating over the velocity distribution, we may decrease the integral boundary from ($-\infty, \infty$) to $[-1, 1]$ by using variable substitution, $\boldsymbol{\beta} = \tan\left(\frac{\pi}{2}t\right)$, \textcolor{black}{where $t$ is a dimensionless parameter facilitating the integration by compactifying the velocity space. Therefore,} 
$$\quad \int_{-\infty}^{+\infty} d{u} f({u}) = \int_{-1}^{+1} d{\beta} \gamma^3 f(\gamma {\beta}) =1.$$
We use brackets {\textcolor{black}{$\langle \ldots \rangle$}} that enclose an arbitrary function $K({u})$ to indicate the average value of that function over {\textcolor{black}{velocity $u$}} \citep{melrose1999relativistic,melrose1999dispersion}, $\langle K\rangle = \int_{-\infty}^{+\infty} d{u} K({u}) f({u}), \quad \int_{-\infty}^{+\infty} d{u} f({u}) =1$.
The $\gamma$ average (see Appendix \ref{sec:B}) obtained by substituting the kappa distribution function leads to 
\begin{equation}\label{eq:16}
\langle{\gamma}\rangle =\int_{-1}^{+1} d({\gamma} {\beta}) {\gamma}({\beta}) f({\beta}) =\int_{-1}^{+1} d{\beta} {\gamma}{\gamma}^3 M[1-N{\beta}^2 {\gamma}^2]^{-\kappa}.
\end{equation} 
To prevent ambiguity arising from multiple values, two constants, $M$ and $N$ are introduced
\begin{equation}
    M =  \left( \pi \theta^2 (\kappa -\tfrac{3}{2})\right)^{-\tfrac{1}{2}}  \frac{\Gamma(\kappa)}{\Gamma(\kappa - \tfrac{1}{2})}, \quad
        N = \tfrac{1}{\kappa -\tfrac{3}{2}\theta^2}.
\end{equation}
$M$ and $N$ depend on $\theta$ and $\kappa$. 
Due to the value of $\gamma$ and {\textcolor{black}{dependence of $\beta$ and $\gamma$ on the power $-\kappa$, which in turn affects how these powers contribute to the computation estimation,}} the integral must be computed numerically for an arbitrary distribution function. {\textcolor{black}{The influence of these dependencies on integration estimation methods is discussed further in Section \ref{NumS}}}.
The kappa distribution function {\textcolor{black}{$g({\beta})$}} can then be expressed in terms of {\textcolor{black}{the velocity}} ${\beta}$ as
  \begin{align*}
	g(\beta) =& \left( \pi \theta^2 (\kappa -\tfrac{3}{2})\right)^{-\tfrac{1}{2}}
	\frac{\Gamma(\kappa)}{\Gamma(\kappa - \tfrac{1}{2})} \left( 1+ \tfrac{1}{\kappa -\tfrac{3}{2}} \frac{(\gamma \beta)^2}{\theta^2} \right)^{-\kappa}.\\
\end{align*}       
Using parameters $M$ and $N$, the distribution function can be simplified to
\textcolor{black}{
\begin{equation}\label{eq:16b}
g({\beta}) = M \left[ 1+ N {\beta}^2 (1-{\beta}^2)^{-1} \right]^{-\kappa}.
\end{equation} }
The derivative of $g({\beta})$ with respect to ${\beta}$ is calculated as follows:

\textcolor{black}{
\begin{align}
    \frac{dg({\beta})}{d{\beta}} =& - \frac{\kappa N M 2 {\beta}}{(1- {\beta}^2)^{2}} \left(1 + \frac{N{\beta}^2}{(1 - {\beta}^2)}\right)^{-\kappa -1},
\end{align} }

or as  \textcolor{black}{
\begin{equation}  \label{eq:18}  
	\frac{dg({\beta})}{d {\beta}} =
	-2\kappa N M {\beta} \gamma^{4} \left( 1+N{\beta}^2 \gamma^{2} \right)^{-\kappa-1} .
\end{equation}}

\subsubsection{Maxwell-Jüttner distribution function}
The Maxwell-Jüttner distribution function \textcolor{black}{serves as a benchmark, allowing us to compare with previous studies and validate our methods},
\begin{equation}\label{eq:19}
	g({u}) = \frac{ e^{(-\rho {\gamma})}}{2 K_1(\rho)};\quad \int_{-\infty}^{+\infty} d{u}\, g({u}) =1.
\end{equation}
The above equation gives the distribution function $g({u})$ in terms of the inverse temperature, $\rho = mc^2/k_BT$, where $m$ is the mass of the particle and $k_B$ is the Boltzmann constant.

The distribution in particle velocities is described by a specific equation showing the distribution function $g(\beta)$\citep{2019JPlPh..85c9005R}

\begin{eqnarray}\label{eq:20}
        g({\beta}) &= \frac{\mathrm{e}^{(- \rho{\gamma})}}{2 K_n(1, \rho)}, \\
        \frac{dg({\beta})}{d{\beta}} &= -\frac{\rho{\beta}(1-{\beta}^2)^{\frac{1}{2}} \mathrm{e}^{(-\rho {\gamma})}}{2K_n(1, \rho)}.
\end{eqnarray}
The $K_n(1,\rho)$ is the {\textcolor{black}{McDonald function, is a modified Bessel function of the second type. Its definition can be found in the Appendix~\ref{sec:B}}}.

\subsubsection{Oblique propagation of $W(z)$}
Oblique propagation of waves occurs when the wave vector is not parallel to the background magnetic field. In the {\textcolor{black}{neutron star}} magnetosphere, oblique propagation of waves can lead to the formation of resonant modes, also referred to as Langmuir waves or plasma waves. {\textcolor{black}{ The modes were found, e.g., by}} \citet{rafat2019wave}
\begin{align}\label{eq:24}
	{\beta}_{A}^{2} &= \frac{\Omega_{e}^{2}}{{\omega}_{p}^{2}\langle {\gamma} \rangle}; \quad \Delta {\beta}^2=\frac{\langle{\gamma} {\beta}^2\rangle}{\langle {\gamma} \rangle} =1-\frac{\langle{\gamma}^{-1}\rangle}{\langle {\gamma} \rangle};\notag\\
	{z}_{A}^{2}&= \frac{{b}}{{a}} = \frac{{\beta}_{A}^{2}-\Delta {\beta}^2}{1+{\beta}_{A}^{2}}; \quad {\gamma}_{A}^{2}= \frac{1}{1-{z}_{A}^{2}} = \frac{1+{\beta}_{A}^{2}}{1-\Delta{\beta}^2};
\end{align}
Using Eq.(\ref{eq:2}) and relations ${a} = 1+{1}\diagup{{\beta}_{A}^{2}}$ and ${b} = 1- {\Delta{\beta}^2}\diagup{{\beta}_{A}^{2}}$,
where ${\beta}_A$ represents the normalized Alfvén speed, $\Delta {\beta}^2$ is a quantity related to the {\textcolor{black}{average velocity}} component of the magnetic field, ${z}_A$ is the relativistic Alfvén speed \citep{2019JPlPh..85c9005R} and {\textcolor{black}{$\Omega_{e} = e{B}/m$ is the electron cyclotron frequency.}}

{\textcolor{black}{The angle $\alpha$}} appears in the oblique plasma response tensor solution because it considers the effect of the magnetic field on the behavior of the wave in the plasma, {\textcolor{black}{which is dependent on the angle between the wave vector and the magnetic field.}}
\begin{align}\label{eq:25}
	&\left({a}-\tfrac{{b}}{{z}^2}\right) \left(1-\tfrac{\omega_{p}^{2}}{\omega^{2}}{z}^2 W({z})-\frac{{b}\tan^2\alpha}{{z}^2} \right)-\frac{{b}\tan^2\alpha}{{z}^2} =0,\\
	&\omega^2 = \omega_{p}^2 {z}^2 W({z}) \left[-\left({a}-\tfrac{{b}}{{z}^2}\right)^{-1} \left(\frac{{b}\tan^2\alpha}{{z}^2}\right)^2 -\frac{{b}\tan^2\alpha}{{z}^2} +1 \right].\notag
\end{align}
The dispersion equation governing the behavior of waves propagating at the angle $\alpha$ can be expressed as follows\citep{melrose1999dispersion}
\begin{equation}
	\frac{\omega^2}{\omega_{p}^2} = \frac{{z}^2W({z})({z}^2-{z}_{A}^2)}{{z}^2-{z}_{A}^2-{b}\tan ^2 \alpha}.
\end{equation}

\subsubsection{The slightly oblique (cold plasma limit) }
The dispersion relation for the slightly oblique wave propagation in the cold plasma limit is characterized by small $\alpha$ angles, which are close to the parallel mode but not strictly aligned. This description corresponds to the cold plasma limit for such configurations, where deviations from strictly parallel propagation can still be analyzed systematically.  The slightly oblique mode is characterized by the specific limits of different parameters. When the inverse temperature $\rho$ approaches infinity, the average Lorentz factor of the particles $\langle{\gamma} \rangle$ approaches one, and the following values are obtained for the parameters above
${a}=1+ \frac{1}{{\beta}^2_{A}}$, ${b}=1$, and ${z}^2 W({z})=1 $.

In the case of an infinite magnetic field strength, where the electron cyclotron frequency ${\Omega}_e$ and Alfvén velocity ${\beta}_A$ approach infinity, the values of the parameters from Eq.(\ref{eq:2}) change. The dispersion relation can be as follows
\begin{equation}\label{eq:26}
	{\Lambda}_{11}={a}-\tfrac{1}{{z}^2}, \quad {\Lambda}_{33}=1-\tfrac{\omega_{p}^2}{\omega^2}-\tfrac{\tan ^2 \alpha}{{z}^2}, \quad{\Lambda}_{13}=\tfrac{\tan \alpha}{{z}^2}
\end{equation}
with the constraints ${a=1, b = 1, z_{A}^2 = 1}$.

By solving a set of Eqs.(\ref{eq:26}), we can obtain the expression for the frequency $\omega$, which is given by
\begin{align}\label{eq:27}
	& \left({a}-\tfrac{1}{{z}^2}\right) \left(1-\tfrac{\omega_{p}^{2}}{\omega^{2}}-\frac{\tan^2\alpha}{{z}^2}\right)-\left(\frac{\tan^2\alpha}{{z}^2}\right)^2 =0,\\\notag
	&\omega^2 = \omega_{p}^2 \left[1-\frac{\tan^2\alpha}{{z}^2}-\left(\frac{\tan^2\alpha}{{z}^2}\right)^2 \left({a}-\tfrac{1}{{z}^2}\right)^{-1} \right]^{-1}.
\end{align}
The Eq.(\ref{eq:25}) is simplified to the specific case of slightly oblique waves in cold plasma limit. 
Finally, the crossover frequency $\omega_{co}$, which evolves the wave characteristics in the plasma and denotes a change in how the plasma responds or behaves at that frequency, can be calculated as $\omega_{co}=\omega_p[{z}_{A}^2 W({z}_A)]^{1/2}$, where $W({z}_A)$ is the plasma dispersion function evaluated at point ${z}_A=\omega/\sqrt{2}\Omega_e$.

\subsection{Numerical solution}\label{NumS}

We developed a one-dimensional Python code that focuses on Langmuir-like waves in a linear Vlasov dispersion solver {\textcolor{black}{of an arbitrary relativistic one-dimensional distribution function}}. We employed rigorous mathematical parameters for integration's high precision. The foundation of our work lies in the solution of the response tensor equation [Eq.(\ref{eq:15})]{\textcolor{black}{. We focus on implementing}} of Eqs.(\ref{eq:20}) and (\ref{eq:21}) for Maxwell-Jüttner distribution and Eqs.(\ref{eq:16b}) and (\ref{eq:18}) for kappa distribution, as nested Python modules.

We designed a module for the numerical solution of the dispersion relation. This module is based on the distribution functions and encompasses the call to parameters such as the phase velocity, kappa index, and temperature parameters for the kappa distribution function. For the Maxwell-Jüttner distribution function, we consider the phase velocity and inverse temperature parameters. \textcolor{black}{By taking advantage} of a quadrature integrator we separate distribution functions in the cases where $|{z}| > 1$ and $|{z}| < 1$.
{\textcolor{black}{The numerical integration technique from the SciPy package, facilitated the solution of the Cauchy integrals, which was achieved using Eq.(\ref{eq:15}) and quadrature integration method, considering the parameters and accuracy levels mentioned above}}.

To ensure the highest integration precision, as it is tunable in quadrature integration with an arbitrarily small number $\epsilon$ equal to $10^{-7}$ in the denominator of Eq.(\ref{eq:4}), we set an upper bound of subintervals as 20 000 and absolute and relative errors of $10^{-10}$. In addition, we applied numerical integration to the derivative of the dispersion function, specifically employing the trapezoidal method, which yields an accuracy of 0.001 for the permittivity derivative.

In addition to the aforementioned modules, we developed a separate module for oblique waves. These waves require different dispersion equations compared to the longitudinal mode.

We used a modified root-finding algorithm which {\textcolor{black}{ searches for a range of wave numbers}}. Motivated by the research of \citet{manthei2021refining}, we {\textcolor{black}{represent}} a significant understanding of the frequency related to the propagation angles and the appropriate wave number. 

Levenberg-Marquardt algorithm, a nonlinear least squares algorithm first proposed by \citet{levenberg1944algorithm} and later developed by \citet{marquardt1963algorithm}, is also incorporated into the method. This algorithm, integrated with the multidimensional root-finding function 'root' from Python’s SciPy library, {\textcolor{black}{finds}} possible solutions and provides separate outcomes for complex and non-defined variables. The ability to define nested functions that impose constraints on the calculations is also made possible, which is especially helpful to incorporate conditions such as light line limitation and eliminating undefined variables.

We utilize the numerical Vlasov dispersion solver algorithm to study Cauchy integrals with sufficiently high precision to avoid errors associated with simpler methods of numerical integration. The Vlasov approach employs differentiation, integration, and power series tools to study the dielectric function and describe real and imaginary parts of frequencies in relation to the wave numbers. The method uses numerical derivatives by applying the trapezoid method to identify the zeros of the permittivity function and depends on defined frequency values obtained from the root-finding algorithm. \textcolor{black}{These frequency solutions are crucial for analysis, as they satisfy the dispersion relation and fall within the range of interest, \( 0.01 < \frac{\omega}{\omega_p} < 5 \), which encompasses frequencies near the plasma frequency. For frequencies approaching zero, the behavior remains consistent, while for frequencies exceeding 5, the differences become negligible, reinforcing that our focus is on the region around the electron plasma frequency.}

The choice of parameters, such as the Alfvén velocity ${\beta}_{a}$ and the plasma frequency ($\omega_p$), both independent of the Lorentz factor, plays a crucial role in this study. These \textcolor{black}{contribute to a consistent structure} in the dispersion relation. In dispersion relation Eq.(\ref{eq:24}), ${z}^2_{a}$ can be approximated as $ \frac{{\beta}^2_{a}}{1 + {\beta}^2_{a}}$. This \textcolor{black}{relation ensures a uniform scaling in the dispersion properties.}

Our method of searching for dispersion tensor solutions is different from the method used by \citet{manthei2021refining}. We search for numerous solutions in frequency for each $k$, whereas they first found one dispersion solution and then tracked the solution in the $(\omega,{k})$ space. Our approach allows to find all present dispersion branches, including those that are hard to find by other approaches. 
 
\section{RESULTS}
\subsection{Differences between kappa and Maxwell-J\"{u}ttner distributions}
\begin{figure*}
	\centering
	\includegraphics[width=\textwidth]{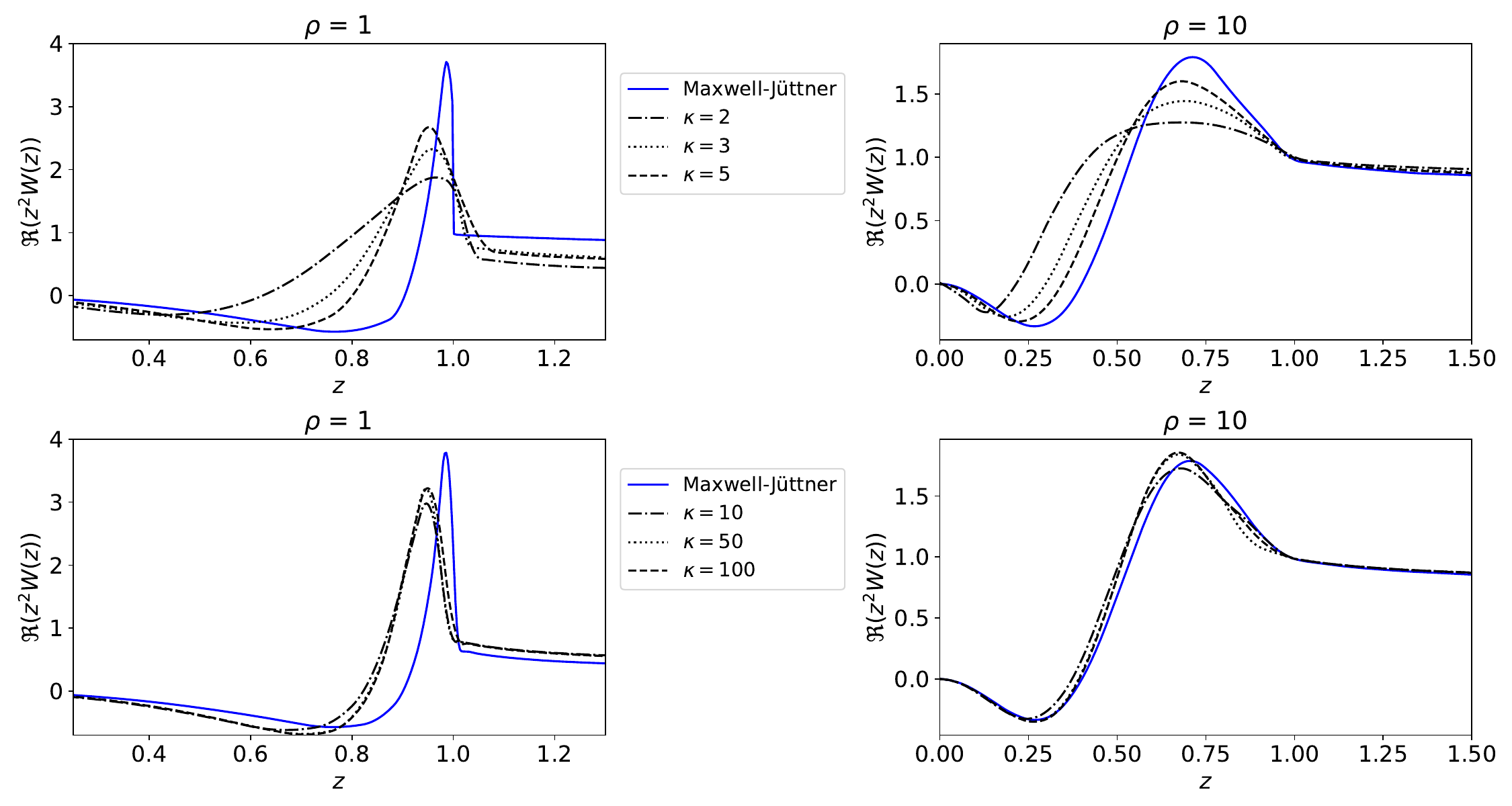}
	\caption{A comparison between the real part of the ${z}^2W({z})$ function for different kappa values and Maxwell-Jüttner distribution functions at $\rho=1$ and $\rho=10$. The ${z}^2W({z})$ function shows flattered behavior where phase velocity \textcolor{black}{approaches the speed of light}.}
	\label{fig:RKdis}
\end{figure*}

We compare the real solutions of the kappa and Maxwell-Jüttner distribution function response tensors. Taking $z=1$ as the critical point and specifically focusing on the $|z|<1$ and $|z|>1$ intervals for inverse temperatures {\textcolor{black}{(where $\rho = mc^2/T$, for example for $\rho=1$ the $\theta$ value corresponds to $0.6\times10^{10}K$ and from now on we denote the temperature by $\rho$ value)}}, $\rho = 1$ and $\rho = 10$, Fig.\ref{fig:RKdis} shows the component of dispersion's response tensor, ${z}^2W({z})$. \textcolor{black}{We study a range from a relativistic temperature $\rho=1$, a semi-relativistic temperature $\rho=10$ to a non relativistic temperature $\rho=100$}

\textcolor{black}{Considering fixed temperature $\rho=1$, we emphasis on how variations in $\kappa$-index influence the distribution function profile. The kappa distribution displays a prominent peak characterized by significant variations in its bulk and tails in the range ${z}\sim(0.5,1.0)$, reaching a maximum value of ${z}^2W({z})\approx 2.8$ for $\rho=1$ and $\kappa=5$. Specifically, for $\rho=1$ the bulk of the distribution function mainly shapes the behavior of ${z}^2W({z})$. In contrast, the central region $0<{z}<0.5$ of the distribution profile is weakly influenced by the variation of $\kappa$}.   

The Maxwell-Jüttner distribution profile has a distinguished behavior from the kappa distribution. For ${z}<1$, the dispersion \textcolor{black}{has} narrower and higher peaks. Whereas, a broad crest in the negative values of ${z}^2W({z})$ suggests a wider core of the distribution function. The plateau line for ${z}>1$ indicates only a weak interaction with the subluminal part of the distribution function.

\begin{figure*}
	\centering
	\includegraphics[width=\textwidth]{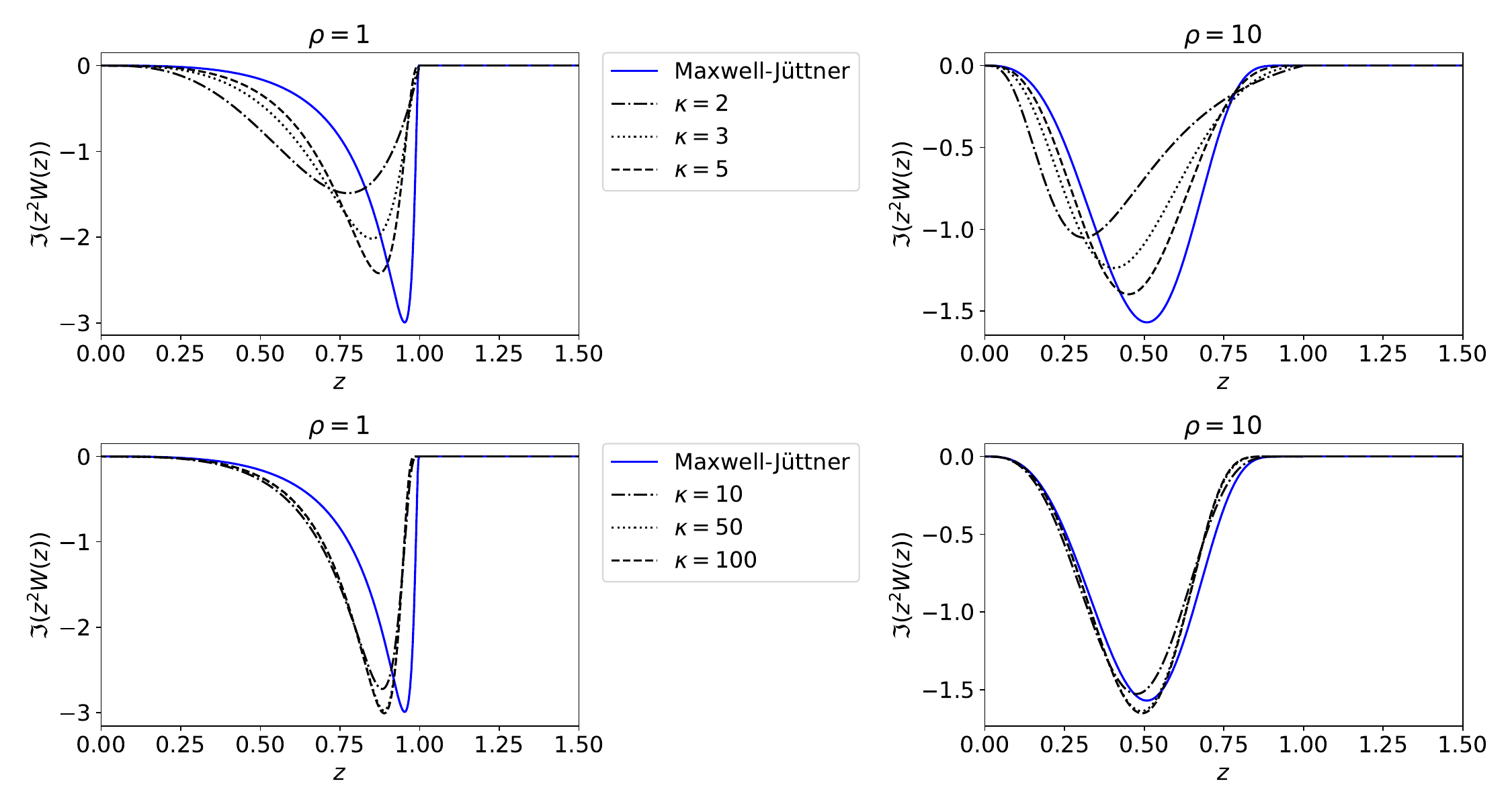}
	\caption{A comparison between the imaginary part of the ${z}^2W({z})$ function of the kappa and Maxwell-Jüttner distribution functions at inverse temperature parameters $\rho=1$ and $\rho=10$ highlights negative peak \textcolor{black}{for both profiles}.}
	\label{fig:IKdis}
\end{figure*}

Figure.\ref{fig:IKdis}, \textcolor{black}{shows} the imaginary \textcolor{black}{part} of ${z}^2W({z})$ under both relativistic and non-relativistic inverse temperatures, $\rho=1$ and $\rho=10$.
For $\rho=10$ and lower $\kappa$-indices $(\kappa=2, 3, 5)$,\textcolor{black}{the kappa distribution  profile resembles a plateau at ${z}\sim 0.3-0.5$ that extends to ${z}^2W({z})\sim0.7$ displaying a wide trough that gradually levels out to a plateau for ${z}>1$, while the Maxwell-Jüttner distribution profile declines gently, reaching ${z}^2W({z})\sim-1.5$, before transitioning sharply into a plateau for ${z}>1$.}

    For $\rho=1$, the minimum profile \textcolor{black}{decreases} away from the center (${z}=0$) plateau. For all kappa values, the profile \textcolor{black}{trough} remained wide, with no distinct trend at $|{z}|>1$. The Maxwell-Jüttner \textcolor{black}{dispersion profile} exhibits an entirely negative profile near \textcolor{black}{$\Im({z}^2W({z}))\sim-3.0$} and displays a sharper trough.

\begin{figure*}
	\centering
	\includegraphics[width=\textwidth]{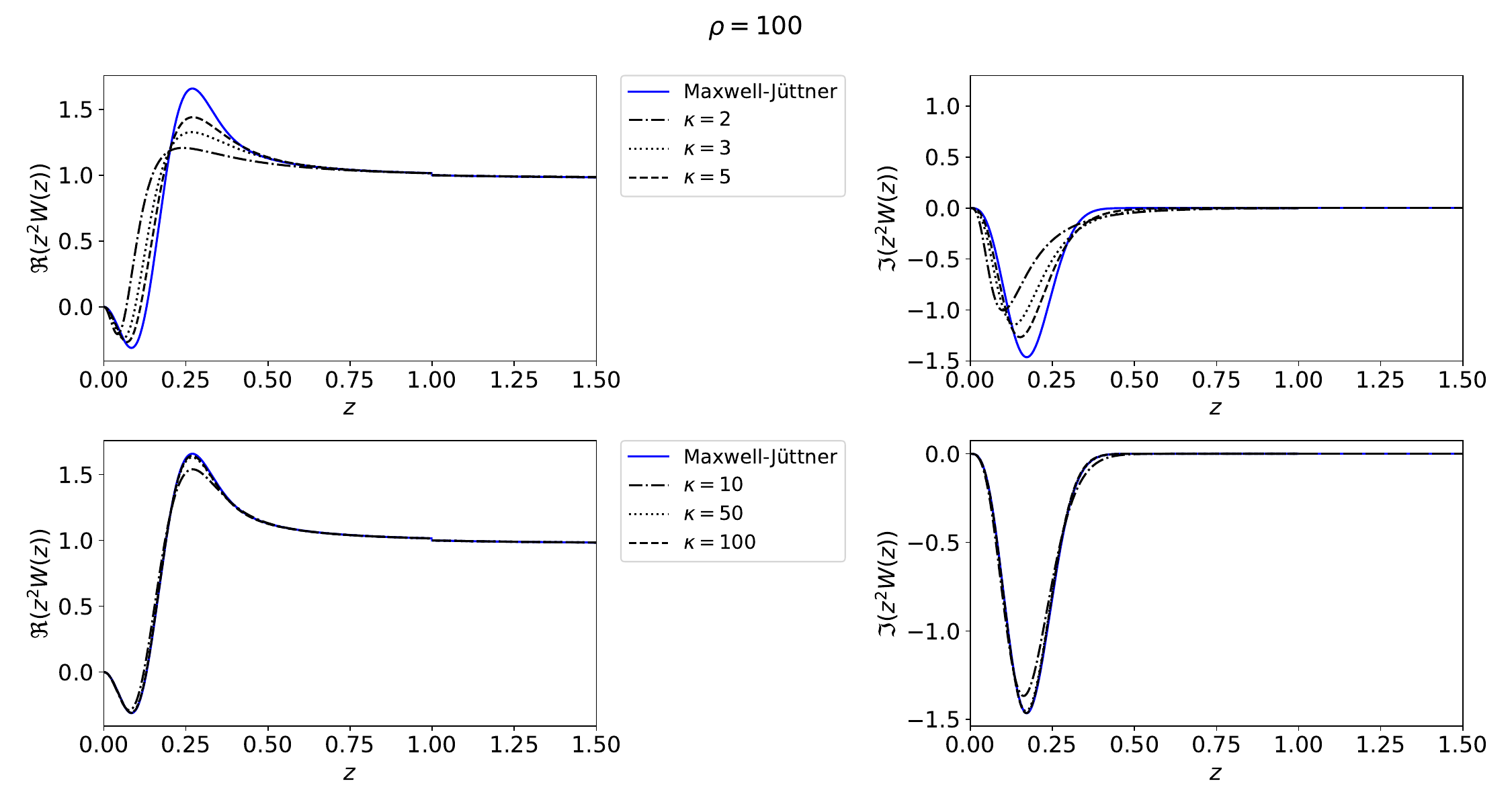}
\caption{Behavior of the ${z}^2W({z})$ functions for low temperature $\rho=100$. The profiles of the lower $\kappa$-indices exhibit peaks and troughs with a narrow range of phase velocities. In the real part, higher $\kappa$-indices approximately overlap with the Maxwell-Jüttner distribution and the imaginary parts \textcolor{black}{show resembling behavior}.}
	\label{fig:RIKdis}
\end{figure*}

\textcolor{black}{For non relativistic temperatures} as $\rho$ approaches 100, in Fig.\ref{fig:RIKdis}, a transition to a Maxwell-Jüttner distribution demonstrates in the behavior of ${z}^2W({z})$. The kappa distribution aligns more closely with the Maxwellian distribution, and a low peak appears at the center for the real components of ${z}^2W({z})$. Lower $\kappa$-indices exhibit slight variations in peak profiles, with $\kappa=5$ being the closest to the Maxwell-Jüttner distribution function. All profiles share tails for ${z}>1$ and exhibit plateau distributions at higher phase velocities. The higher $\kappa$-indices align with Maxwell-Jüttner in the real components of the ${z}^2W({z})$ response tensor. In the imaginary part of the response tensor, all kappa profiles have \textcolor{black}{negative troughs same as Maxwell-Jüttner. Higher $\kappa$-indices exhibit troughs far from the center and $\kappa=2$ has a wider trough}. Compared with relativistic temperatures, the imaginary components of the response tensor in lower temperatures show narrower peaks.


\subsection{Dispersion relations}
\begin{figure}[t]
	\centering
   \includegraphics[width=\columnwidth]{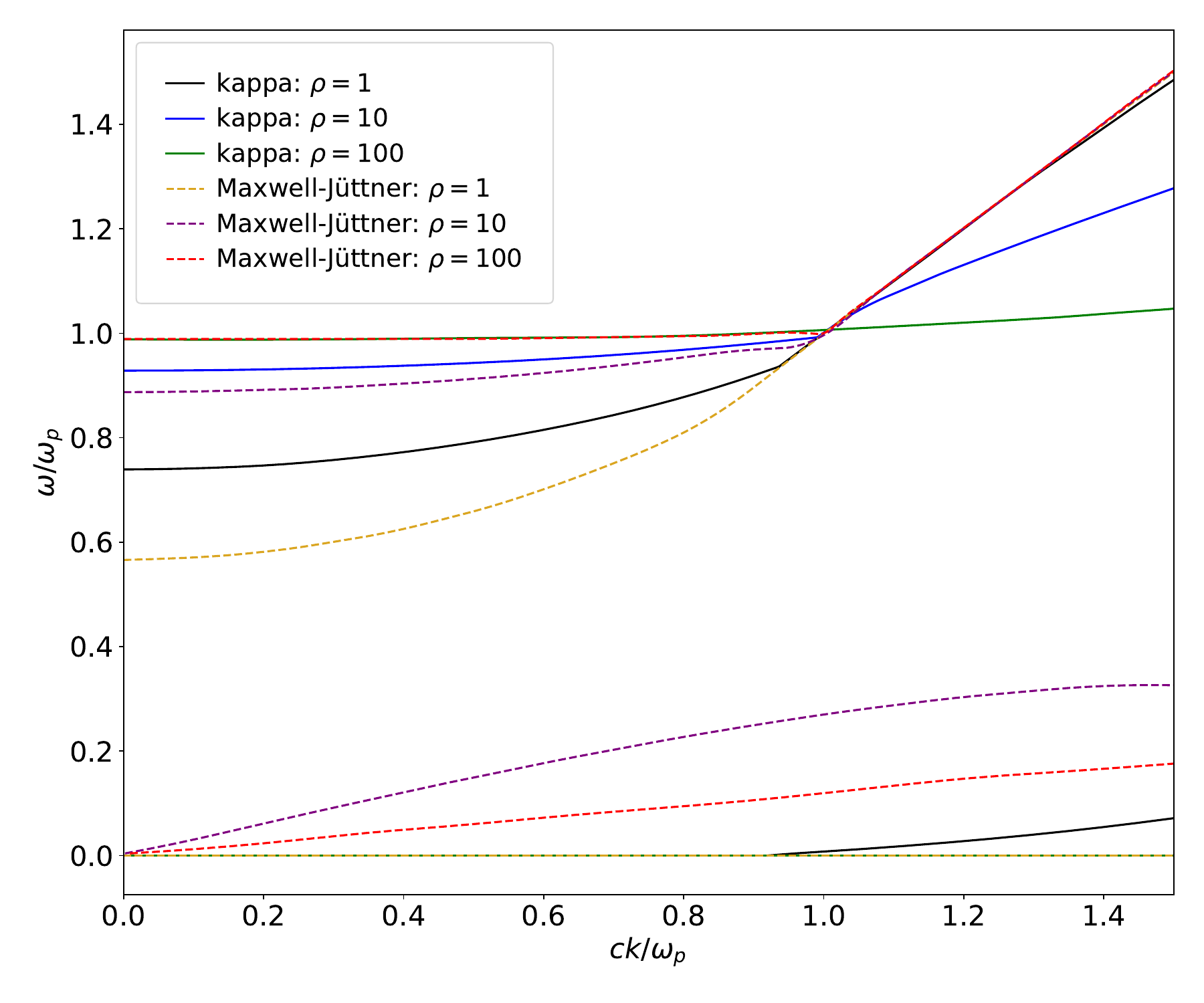}
	\caption{The dispersion branches of the kappa and Maxwell-Jüttner distributions, showing dispersion branches for each temperature. Dashed lines represent Maxwell-Jüttner distribution and solid lines represent kappa distribution dispersion branches.}
	\label{fig:AMJK21}
\end{figure}

For all subsequent calculations, the $\kappa$-index is equal to 2, because $\kappa=2$ is the most distinct version of the Maxwell-Jüttner profiles in the relativistic regime.
Figure.\ref{fig:AMJK21} displays the dispersion branches of the kappa distribution and the Maxwell-Jüttner distribution by comparing the wave properties of the thermal plasma, from relativistic to non-relativistic temperatures. The dispersion profiles show two dispersion branches for each wave number with lower branches {\textcolor{black}{indicating}} higher temperatures.

In addition, the dispersion branches of the kappa distribution \textcolor{black}{indicate} that the upper branch \textcolor{black}{shows an upward trend} for all temperatures in terms of frequency, while the lower branches go straight except for the $\rho=1$ branch. The lower branches start at $\omega=0, k=0$.

Because the dispersion branches for slightly oblique waves show only small differences from the parallel case, we plot the oblique modes to highlight the differences. Figure.\ref{fig:AKO} represents the oblique modes of the kappa \textcolor{black}{distribution}. In particular, at zero frequency, \textcolor{black}{the plot shows no oscillation and the wave is either heavily damped or not propagating}.
At higher temperatures, $\rho=1$, a scattered response \textcolor{black}{of the particles to the magnetic field occurs} at a zero angle \textcolor{black}{or the parallel mode} with respect to the magnetic field. The deviation from $\alpha=0$ \textcolor{black}{branches} increases with increasing angle.

In addition, the figure illustrates that the branches deviate \textcolor{black}{and curve up} more with higher temperatures, showing similarities between $\rho=10$ and $\rho=100$ at $\alpha=0.5, 0.25$. The result also demonstrates the impact of temperature on the dispersion curves, resulting in straighter diagonal branches of $\rho=1$. The upper solid branch is slightly curved up for ${k}c/\omega_p<1$, then increases sharply in frequency \textcolor{black}{for lower inverse temperatures. But} for higher inverse temperatures, it is \textcolor{black}{almost} horizontal for all ${k}c/\omega_p$.

\subsection{The permittivity derivative}
\begin{figure}[t]
	\centering
	\includegraphics[width=\columnwidth]{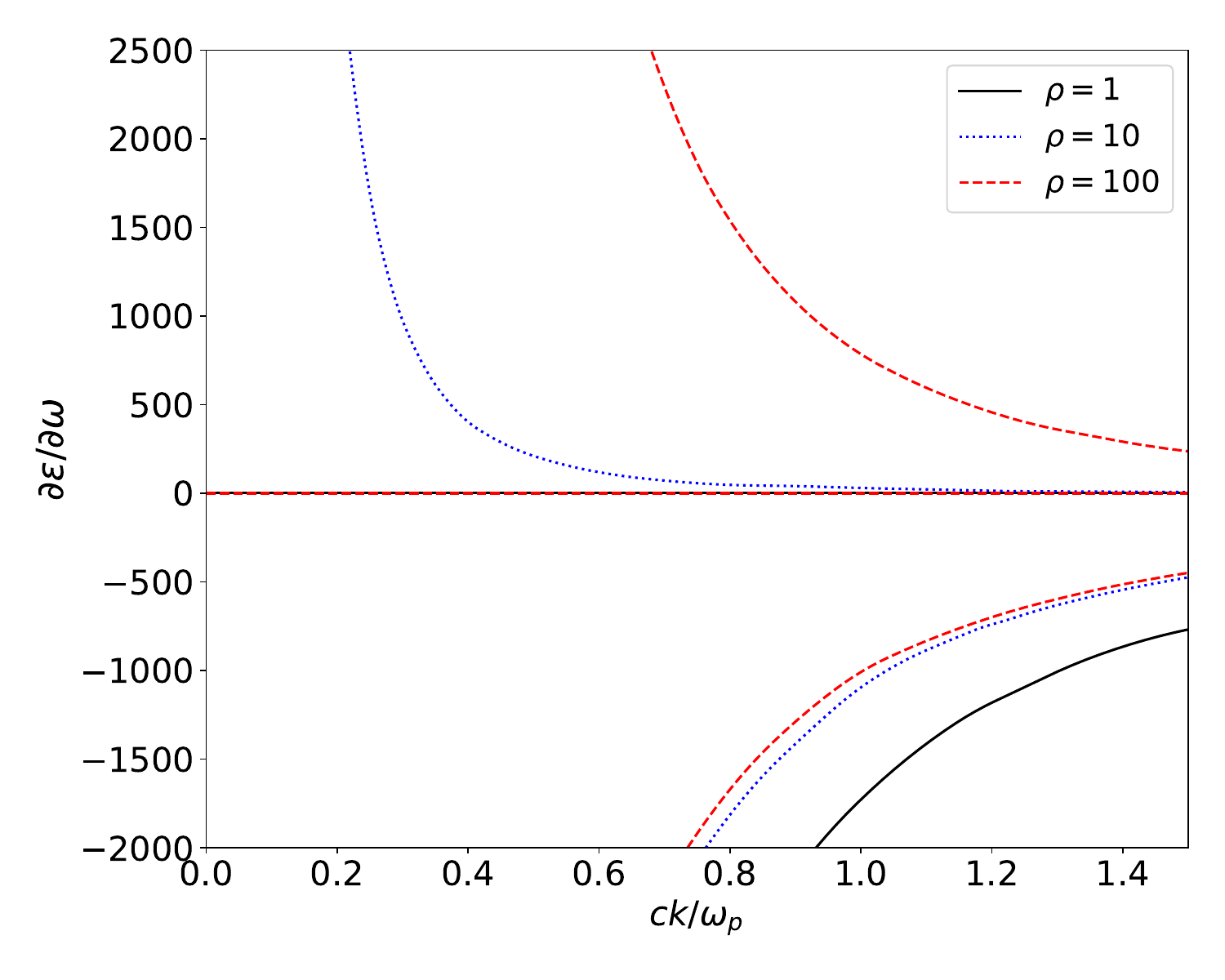}
	\caption{The permittivity derivative profile of the Maxwell-Jüttner distribution demonstrates a significant increase in the negative half of the frequencies and a decrease in the positive half of the frequencies.}
	\label{fig:BMJn}
\end{figure}

Figure. \ref{fig:BMJn} shows two sets of branches with positive and negative derivatives, both approaching zero by increasing the wave number for the Maxwell-Jüttner distribution.
The increasing derivative values with decreasing temperature suggest a more diverse behavior compared to the decreasing branches.
As a result of decreasing the inverse temperature and dispersion branches approaching the light line, the permittivity derivative branches end at zero, \textcolor{black} {which suggests the tendency to stable propagation or the resonance condition. This interpretation matches the ${z}^2W({z})$ profile in Figs. \ref{fig:RKdis}--\ref{fig:RIKdis}.}  

\begin{figure*}[t]
	\centering
	\includegraphics[width=\textwidth]{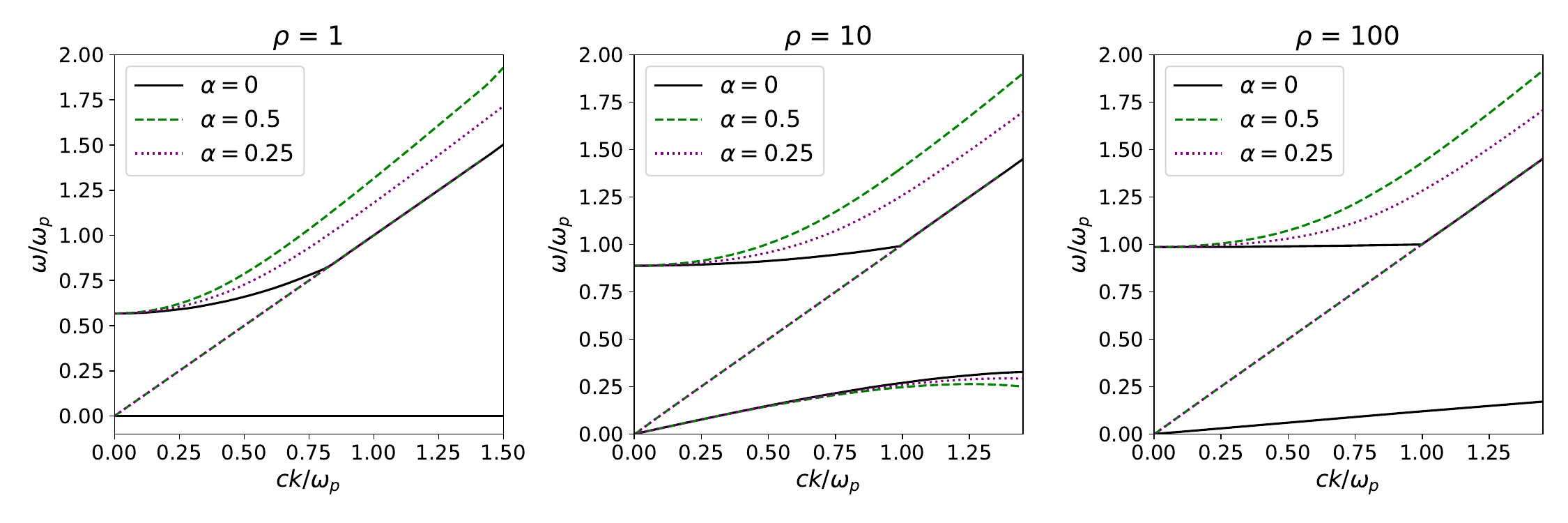}
	\caption{The Maxwell-Jüttner distribution's  dispersion branches for oblique modes as a function of the inverse temperature. The solid line approaches the light line while the dashed lines deviate towards higher phase velocities.}
	\label{fig:AMJO}
\end{figure*}

\begin{figure*}
	\centering
	\includegraphics[width=\textwidth]{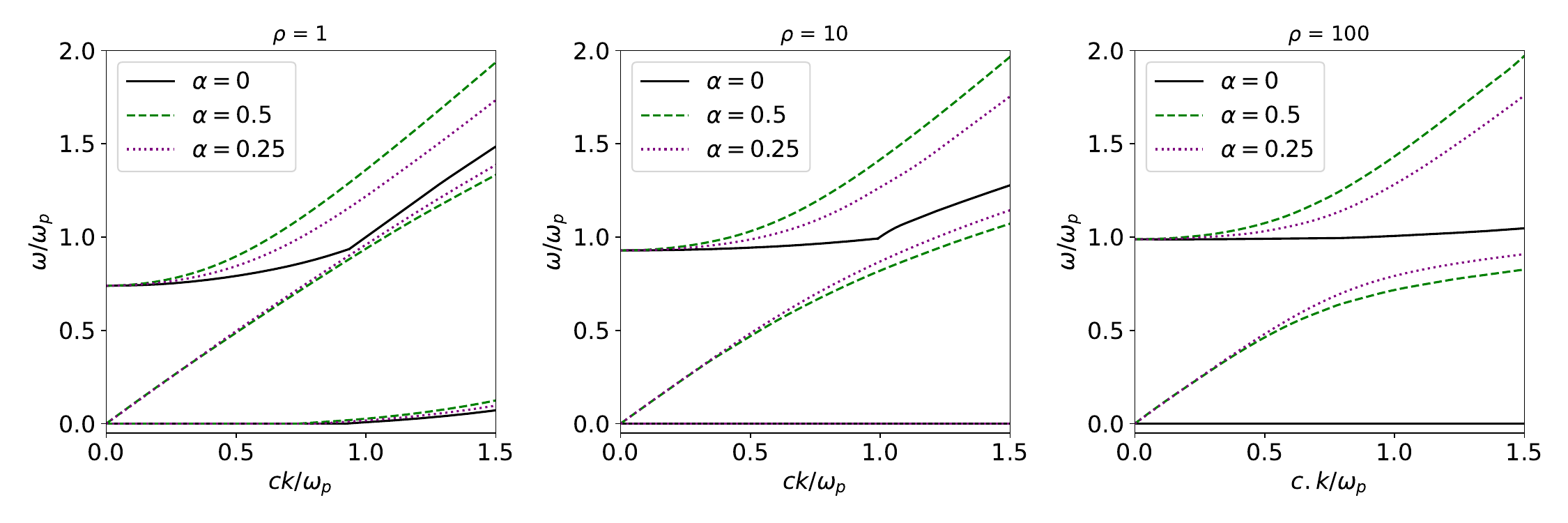}
	\caption{The kappa distribution's dispersion branches for oblique modes demonstrates the similar effect as in the Fig. \ref{fig:AMJO}.}
	\label{fig:AKO}
\end{figure*}

\section{DISCUSSION}
To analyze the {\textcolor{black}{dispersion properties of neutron star }} magnetospheric plasma, we {\textcolor{black}{developed a numerical tool to analyze arbitrary distribution functions, looking}} at the relativistic kappa and Maxwell-Jüttner distribution functions. {\textcolor{black}{We considered}} thermal effects in strongly magnetized neutron star magnetospheric plasmas and compared their real solutions \textcolor{black}{with each other for relativistic to non-relativistic plasmas}. 

\subsection{Distribution functions}
Comparison of the kappa and Maxwell-Jüttner distribution functions \textcolor{black}{provides insight into the velocity distribution characteristics in plasmas.} The kappa distribution revealed slightly lower profile peaks of ${z}^2W({z})$ at relativistic temperatures, while at lower temperatures, the Maxwell-Jüttner and kappa distribution profiles show \textcolor{black}{matching response tensor profiles for real solutions in the dispersion relation.} ${z}^2W({z})$ for both distribution functions, particularly in the profile center, \textcolor{black}{show a stable} velocity distribution\textcolor{black}{, where is no net transfer of particles between different velocity ranges}. The profile ${z}^2W({z})$ shows a plateau line for ${z}>1$. For imaginary part of \textcolor{black}{${z}^2W({z})$} \textcolor{black}{demonstrated} in Fig.\ref{fig:IKdis}, the Maxwell-Jüttner distribution \textcolor{black}{presents} sharper negative peaks at relativistic temperatures. \textcolor{black}{Similarly, the kappa distribution features negative peaks at both non-relativistic and relativistic temperatures, but with a broader profile compared to the Maxwell-Jüttner distribution.}

Figure.\ref{fig:RIKdis} indicates that at $\rho=100$ as a non-relativistic temperature, the $\kappa$-distribution function for higher $\kappa$-indices approaches the Maxwellian distribution; the behavior of the ${z}^2W({z})$ profile shows a fluctuating pattern \textcolor{black}{for ${z} = 0 - 0.1$}. We found that the choice of $\kappa$-index can significantly impact the $z^2W(z)$ function and \textcolor{black}{gradually decrease its amplitude}.

\textcolor{black}{As $\kappa$ increases, the ${z}^2W({z})$ profiles converge to those of $\kappa = 100$, reflecting the kappa distribution’s approach to its thermal limit. This trend indicates that the plasma transitions to a state of local thermodynamic equilibrium, with nonthermal effects becoming negligible. In relativistic neutron star plasmas, higher $\kappa$ values signify a dominance of thermal particles, reducing the impact of high-energy processes and aligning the system closely, though not identically, with the Maxwell-Jüttner distribution.}

\subsection{The dispersion relation}
As shown in Fig.\ref{fig:AMJK21}, the dispersion branches of the Maxwell-Jüttner and the kappa distribution functions exhibit relativistic characteristics at higher temperatures, as \textcolor{black}{indicated by upward-curving dispersion branch for $\rho=1$}. The straight lines in the dispersion branches at lower temperatures indicate a \textcolor{black}{stable frequency of the dispersion branch} in response to the magnetic field. In addition, a straight dispersion branch indicates waves propagating with a relatively constant \textcolor{black}{group} velocity over a wide range of frequencies and wave numbers.

The \textcolor{black}{other curved trends of diagonal lines} match the results obtained by \citet{manthei2021refining}, where they compared the analytical modeling of the subluminal L-mode branch with the result calculated by the PIC model. Therefore, our relativistic wave modes are compatible with the method utilized by \citet{manthei2021refining}.

Figure.\ref{fig:AMJK21} highlights the response of the {\textcolor{black}{neutron star}}'s magnetospheric plasma to the magnetic field at lower temperatures, characterized by \textcolor{black}{a smoother behavior in the upper branch for both profiles compared to higher-temperature cases, which is more pronounced in the kappa distribution. Meanwhile, a stronger deviation from the corresponding upper branch trend in the lower branch is shown in the Maxwell-Jüttner profile.}  

In general, additional branches may arise within narrow frequency ranges. It appears that some of these can be classified as weak phenomena with limited energy-carrying capacity, as our root-search solver did not detect them completely. This finding aligns with the assertions made in \citet{benavcek2019growth}, where it is noted that such weak branches may emerge \textcolor{black}{for some wave numbers}, while nearly disappearing for \textcolor{black}{different ones}.

\subsection{Dispersion properties of the oblique modes}

\textcolor{black}{Our findings indicate that the distribution function has only a subtle influence on dispersion properties at small angles $(\alpha \to 0)$, although it may play a more significant role under different conditions or larger obliquities. This work focuses on a wide range of propagation angles, excluding slightly oblique cases where no significant results were observed. By testing this framework with Maxwell-Jüttner and kappa distributions, we establish its reliability for future exploration of more realistic and intricate distribution functions}. In Fig.\ref{fig:AMJO}, we investigate the dispersion branches of oblique modes \textcolor{black}{for} the Maxwell-Jüttner distribution \textcolor{black}{for varying angles and temperatures}. Our results show that \textcolor{black}{as} the inverse temperature decreases, the \textcolor{black}{frequency} deviation of the dispersion branches \textcolor{black}{from the parallel-propagating mode increases}.

 Furthermore, it is shown that in the oblique mode, \textcolor{black}{the dispersion branches increasingly diverge from those at zero angle as the angle $\alpha$ deviates further from the direction of the magnetic field}. This divergence from the parallel mode can be attributed to the angle's effect on wave propagation, which modifies the relation between wave number and frequency. The zero angle responds distinctly to the magnetic field at higher temperatures, $\rho=1$.
The dispersion branches of the kappa distribution in Fig.\ref{fig:AKO} shows that deviations from the parallel mode \textcolor{black}{which} appear in the \textcolor{black}{profiles are} less prominent at lower wave numbers and frequencies, \textcolor{black}{in addition, at $\rho = 10$, the profile shows a small numerical artifacts at $z=1$ for parallel propagation $\alpha=0$ at ${z}>1$ visible as slightly curved line.}

\subsection{The permittivity derivative profile}
The Maxwell-Jüttner distribution derivative of permittivity branches converges to zero as the wave number increases, with higher derivative values at lower temperatures indicating more \textcolor{black}{variation in permittivity}. With decreasing \textcolor{black}{the} inverse temperature and dispersion branches close to the light line, the permittivity derivative branches ultimately reach zero. The zero permittivity derivative line indicates an equilibrium phase.

\section{CONCLUSION}
Our research compared the \textcolor{black}{dispersion properties of the magnetospheric neutron star plasmas following the} Maxwell-Jüttner and kappa distribution functions. \textcolor{black}{We found} that the kappa distribution provides a practical description over a broad temperature range, considering different kappa indices. Our analysis shows that the kappa distribution converges to the Maxwell-Jüttner distribution under specific conditions, namely when $\kappa \to \infty$, $\rho \to \infty$, and the plasma temperature approaches zero. Outside these limits, the power-law tail and nonthermal characteristics of the kappa dispersion properties remain significant, differentiating it from the Maxwell-Jüttner distribution. We include this comparison to highlight both the similarities in specific \textcolor{black}{plasma conditions, such as relativistic and semi-relativistic versus thermal cases.}
\textcolor{black}{However, the kappa distribution approaches the Maxwell-Jüttner form, indicating a transition to thermal equilibrium. At high $\kappa$, nonthermal effects diminish, and the plasma becomes dominated by thermal particles.}

Using the NumPy and SciPy libraries, we \textcolor{black}{developed a Python-based linear dispersion solver capable of handling} \textcolor{black}{arbitrary} distribution functions, \textcolor{black}{provides} solutions to the dispersion relation in terms of wave number, frequency, and group velocity. \textcolor{black}{By applying this method, we gain deeper insights into the dispersion properties of plasma waves, including the plasma response to the magnetic field and their behavior as they approach relativistic speed.}

The \textcolor{black}{dispersion} function profiles show similarities at higher wave numbers and lower temperatures, indicating that the analysis in the low-temperature region may not be significantly impacted by the \textcolor{black}{the choice of} distribution function. \textcolor{black}{Additionally, the dispersion properties} tend to converge toward a Maxwellian distribution. However, at lower wave numbers and higher temperatures, each dispersion relation showed \textcolor{black}{distinct} peaks and troughs. This suggests that \textcolor{black}{investigating} the kappa distribution is also useful in obtaining a \textcolor{black}{detailed} understanding of plasma processes within the \textcolor{black}{neutron star} magnetosphere or comparable \textcolor{black}{astrophysical environments with similar characteristics}. \textcolor{black}{The negative imaginary part is consistent with the kappa and Maxwell-Jüttner distributions, which are prevalent in non-equilibrium, collisionless plasmas. In addition, it signifies energy loss or damping, reflecting their role in describing relativistic thermal plasmas.}

Our modeling results indicate that \textcolor{black}{strong electrostatic waves are enhanced in the studied wave number range particularly in regions where the dispersion properties deviate from the Maxwell-Jüttner distribution}.
Comparing the oblique modes of the Maxwell-Jüttner distribution\textcolor{black}{, we find} that the kappa distribution's dispersion branches \textcolor{black}{exhibit} converging behavior \textcolor{black}{as wave angles increase relative to the magnetic field direction, along with variations in inverse temperatures and plasma responses to the magnetic field. In both relativistic and non-relativistic temperature ranges, distinct dispersion characteristics are observed}.

Analyzing the oblique modes of the kappa distribution's dispersion branch \textcolor{black}{reveals deviations from the light line in the non-zero angle profiles, particularly within a specific range of frequencies and wave numbers. These deviations tend to increase, indicating a more pronounced frequency difference from the light line as the inverse temperature parameter increases, suggesting larger modifications in the dispersion properties, for the magnetospheric plasma. In contrast, such angle-dependent variations are not observed for the Maxwell-Jüttner distribution.}

\textcolor{black}{While the influence of the distribution function is subtle at small angles ($\alpha \to 0$), it becomes more significant at larger propagation angles. This highlights the importance of exploring oblique modes in highly relativistic and magnetized environments such as neutron star magnetospheres, where complex plasma distributions cannot be fully captured by simple forms such as Maxwell-Jüttner or kappa distributions. Our framework provides a foundation for studying more realistic distribution functions in future analyses.} 

Based on our findings, we recommend further investigations \textcolor{black}{of} streaming instability, considering oblique conditions and our proposed parameters. Furthermore, a comparative examination of the distribution functions and emissions of Langmuir-like waves, utilizing our computed quantities, would offer a significant understanding of the plasma's behavior in the neutron star's magnetosphere.
 
\begin{acknowledgments}
M.M. acknowledges support from the doctoral scholarship within the framework of the scholarship program in the faculty of physics, University of Duisburg-Essen. J.B. acknowledges support from the German Science Foundation (DFG) projects BU 777-17-1 and BE 7886/2-1. The authors gratefully acknowledge the Gauss Center for Supercomputing e.V. (\href{www.gauss-centre.eu}{www.gauss-centre.eu}) for funding this project by providing computing time on the GCS Supercomputer SuperMUC at the Leibniz Supercomputing Center (\href{www.lrz.de}{www.lrz.de}).

\end{acknowledgments}

\appendix

\section{The Landau prescription solution}

The dispersion relation is \textcolor{black}{dependent on} the plasma properties through which the wave propagates. Therefore, the real part of the dispersion relation under \textcolor{black}{the} resonant condition could be written as $1={\omega^2_p}\diagup{{k}^2_{\parallel}}({\beta}-{z})^2$ , where the Langmuir wave frequency matches the plasma frequency, resulting in a resonance between the wave and the plasma particles, \textcolor{black}{which is consistent with the theory in} \citet{chen1984introduction}
\begin{equation}\label{eq:6}
	\lim_{t \rightarrow \infty} \frac{\sin{(\omega-{k}{u})t}}{(\omega-{k}{u})} = \delta(\omega-{k}{u}).
\end{equation}
By considering the lower half-plane as a mathematical construct used in the study of complex functions, one can focus on the singularities of the function, such as zeros or poles, and ignore the residual points. Accordingly, the residual points would be small enough and the roots that have to be zeros, or poles, that are singularities of various types.
On the other hand, if the \textcolor{black}{contour integral for a small damping rate is not applied in the calculation of Landau damping, the following simplified dispersion relation} is considered
\begin{equation}\label{eq:7}
	\dfrac{\omega}{c}={k}_{\parallel}.
\end{equation}

The imaginary part of the frequency $\omega$ plays a critical role in determining the stability of plasma oscillations. In the complex plane of the distribution function, the integral is also related to the damping or decay rate of the plasma oscillations. In general, the imaginary part of $\omega$ should be negative for stable plasma oscillations, implying that the plasma oscillations decay over time. In contrast, a positive imaginary part results in unstable oscillations, leading to the rapid \textcolor{black}{conversion of wave energy into particle kinetic energy}. In other words, the damping of the plasma oscillations is due to the transfer of energy from the oscillating electric field. The frequency $\omega>0$ in the Eq.(\ref{eq:6}) as the resonant denominator, which is related to the sign of ${k}_{\parallel}$, represents the wave vector parallel to the magnetic field. The resonant denominator in Eq.(\ref{eq:6}) is replaced by
\begin{equation}\label{eq:8}
	i\pi sgn({k}_{\parallel})\delta({\beta} - {z}),
\end{equation}
where 
\begin{equation}
	sgn ({k}_{\parallel})=\dfrac{{k}_{\parallel}}{\lvert {k}_{\parallel} \lvert}. 
\end{equation}
To solve the contour integral supposing aforesaid conditions, the singularity is on the contour of complex plane of distribution function therefore $\Im(\omega) = 0$ as stated in \citet{chen1984introduction}
\begin{equation}\label{eq:9}
	\oint d{z} \frac{f({z})}{{z}- {a}} = f({a}) (-\pi i ).
\end{equation}
The principal value solution method as a result from complex analysis, is used for this case of even function in relevant area as the points of the singularity can omit each other, obtaining
\begin{equation}\label{eq:10}
	P.V. \int_{-\infty}^{+\infty} d{u} \frac{f({u})'}{{u}-\tfrac{\omega}{{k}}}+\pi i \frac{df}{d{u}}\mid _{{u}=\tfrac{\omega}{{k}}}.
\end{equation}
Equation.(\ref{eq:10}) relates to the Laplace domain and represents the integral for an initial-value problem. However, as Landau stated, when an initial value problem is considered by performing a Laplace transform, additional contributions exist, as well as by assuming waves in a plasma. An additional term beyond this real part which is an imaginary part, giving some sort of damping that gives a more precise wave-like analysis appears
\begin{equation}\label{eq:11}
	i\pi \frac{{k}_\parallel}{\lvert {k}_\parallel \rvert} \frac{dg({u})}{d{\beta}} \mid_{{\beta}={z}} + \lim_{\delta \rightarrow 0} \left( \int_{-1}^{{z}-\delta} +\int_{{z}+\delta}^{1} \right) d{\beta} \frac{1}{{\beta}-{z}}\frac{dg({u})}{d{\beta}}.
\end{equation}
The function is originally defined by $\omega/{k}$ \textcolor{black}{with $\omega$ as a complex variable and ${u}$ and real as velocity}. To study plasma oscillations, and particularly analyzing the damping and growth of these oscillations, The solution is anticipated to the imaginary part in cases when $\Im(\omega)$ approaches zero. 
The remaining integral can be partially integrated to
\begin{equation}\label{eq:12}
	\lim_{\delta \rightarrow 0} \left( \int_{-1}^{{z}-\delta} +\int_{{z}+\delta}^{1} \right) d{\beta} \frac{1}{{\beta}-{z}}\frac{dg({u})}{d{\beta}}\\
\end{equation}
Altering variables within the integrand to simplify the expression and employing integration by parts, \textcolor{black}{we obtain}
\begin{equation}\label{eq:13}
	\left\{
	\begin{tabular}{@{}c@{}}
		${u} = \frac{1}{{\beta} -{z}} \quad  d{u} = \frac{-2}{({\beta}-{z})^2}$\\
		$dy= dg({u}) \quad v=g({u})$.
	\end{tabular}
	\right.\\
\end{equation}
Equation.(\ref{eq:11}) is a result of partial integration of the integral in Eq.(\ref{eq:10}), which the integrand is simplified by making a change of variables and using integration by parts. The resulting expression is then used to simplify further calculations involving plasma oscillations.
When $\delta \rightarrow 0$ \textcolor{black}{and} using $\quad \vert_{{\beta} = {z}\pm \delta} \thickapprox g({u})\vert_{{\beta} = {z}}$, the result is achieved as well \citep{2019JPlPh..85c9005R}
\begin{equation}\label{eq:14}
	\lim_{\delta \rightarrow 0} \frac{-2g({u})\vert_{{\beta}={z}}}{\delta}+\left(\int_{-1}^{{z}-\delta} +\int_{{z}+\delta}^{1} \right)d{\beta} \frac{g({u})}{({\beta} - {z})^2}.
\end{equation}
Equation.(\ref{eq:12}) is obtained by combining Eq.(\ref{eq:10}) and Eq. (\ref{eq:11}), and taking the limit as $\delta$ approaches zero. The equation describes the behavior of the function near the point ${\beta}={z}$, and it involves the Lorentz factor $\gamma_{\phi} = 1\diagup(1-{\beta}^2)^{-\tfrac{1}{2}},$ which is evaluated at this point. Therefore, at $|{z}| < 1$, the following relation is obtained
\begin{align}\label{eq:15}
	&P.V.\left(\int_{-1}^{1}\frac{g({z})}{({\beta}-{z})^2}d{\beta} \right)\\\notag
	&= \lim_{\delta \rightarrow0^+} \left( \int_{-1}^{{z}-\delta}\frac{1}{({\beta}-{z})^2}d{\beta}+\int_{{z}+\delta}^{1}\frac{1}{({\beta}-{z})^2} d{\beta} \right)\\\notag
	&= \lim_{\delta \rightarrow0^+}g({z})\left(\left(\frac{1}{({\beta}-{z})^2}\right)_{-1}^{{z}-\delta}+ \left(\frac{1}{({\beta}-{z})^2}\right)_{{z}+\delta}^{1}\right)\\\notag
	&=g({z})\left(-2\gamma_{\phi}^2 + \lim_{\delta \rightarrow0^+}\tfrac{2}{\delta} \right).
\end{align}
To further investigate the above mentioned result in the context of plasma oscillations, the phase velocity $\omega/{k}$ is suggested much greater than the typical thermal velocity, implying that the contribution from the integral in Eq.(\ref{eq:10}) for ${u}$ near $\omega/{k}$ is dominated by the residue term, which corresponds to the imaginary part of $dg({u})/d{\beta}$ evaluated at ${\beta}={z}$. This residue term results in damping, and is known as Landau damping in plasma physics. Thus, {\textcolor{black}{the term}} $i\pi dg({u})/d{\beta}$ is the source of the Landau damping in the plasma.

\section{$\langle \gamma \rangle$ evaluation for Maxwell-Jüttner distribution function}\label{sec:B}
{\textcolor{black}{The procedure how obtain the average Lorentz factor of a velocity distribution function can be}} found in the works of  \citet{melrose1999relativistic} and \citet{melrose1999dispersion} as well as those described in the kappa distribution Section \ref{sec:kappa}. The equation for calculating the average ${\gamma}$ is as follows.
$$ \langle {\gamma} \rangle =\int_{-\infty}^{+\infty} d({\gamma} {\beta}) {\gamma}({\beta}) f({\beta}).$$ 
  After extracting $d{\beta}$ from $d{u}$ and substituting Maxwell-Jüttner distribution function the equations turn to
\begin{align}\label{eq:21}
	&d({\gamma} {\beta}) = {\gamma}^3 d{\beta},\\\notag
	&\langle {\gamma} \rangle = \frac{1}{2K_1(\rho)}\int_{-\infty}^{+\infty}{\gamma}^3 {\gamma} \exp{(-\rho {\gamma})}d{\beta}.
\end{align}
Referring to \citet{melrose2008quantum}, the substitution for Lorentz factor is ${\gamma} = \cosh{{x}}$. Here we consider calculations taking into account the McDonald coefficient \citep{melrose1999relativistic}, as the initial value of the McDonald coefficient is $$ 2K_0(\rho) = \int_{-\infty}^{+\infty} \frac{\exp^{i\rho t}}{\sqrt{t^2-1}} ,$$
 while \textcolor{black}{$t$ is a dimensionless variable and the} value of {\textcolor{black}{$\sigma$ order is $$K_{\sigma}(x) = \int_{0}^{+\infty}\exp^{-x\cosh{t}} \cosh{\sigma t}dt .$$}}

Taking advantage of above substitutions can guide one to a substitution as below
\begin{align}\label{eq:22}
	&{\gamma}^3 {\gamma} = \cosh{2{\beta}}  \\\notag
	&\cosh{2{\beta}}^2 = \frac{\cosh{2{\beta}}+1}{2}\\\notag
	&\langle {\gamma} \rangle =\frac{1}{2K_1(\rho)}\int_{-\infty}^{+\infty} \frac{\cosh{2{\beta}}+1}{2}\exp^{-\rho \cosh{{\beta}}} d{\beta}
\end{align}    
Because $\cosh$ is an even function, it exhibits symmetry {\textcolor{black}{in ${\beta}$}} due to its property of maintaining an identical output value when the sign of its independent variable is reversed. Therefore, since $\cosh{(0)}=1$, the {\textcolor{black}{integration range}} can be broken into two integrals
\begin{align}\label{eq:22} 
	&\frac{1}{2K_1\rho}\int_{-\infty}^{0}\cosh{(2{\beta})}e^{-\rho \cosh{({\beta})}} d{\beta} 
	\\&+\int_{0}^{+\infty} \cosh{(0)}e^{-\rho \cosh{({\beta})}} d{\beta}, \nonumber \\
	&\langle {\gamma} \rangle = \frac{K_2\rho+K_0\rho}{2K_1 \rho}.
\end{align}

Finally, to show $g({\beta})$ with respect to ${\beta}$; the equation below shows that $g(\beta)$ is proportional to the product of ${\beta}$, $\exp(-\rho{\gamma})$, and a function of ${\gamma}$. This equation contains a parenthetical expression that involves both ${\beta}$ and ${\gamma}$, which appears as a coefficient in the equation below. 
\begin{align}\label{eq:23}
	&g({\beta}) = {\gamma} (1+ {\beta}^2 {\gamma}^2) \frac{e^{-\rho {\gamma}}} {2K_n(1, r_i)}
	\frac{dg({\beta})}{d{\beta}} \\\notag
	&= \frac{{\beta} e^{-\rho {\gamma}}}{2K_n(1, r_i){\gamma}^3}\left( (3-3{\beta}^3{\gamma}^2)-\rho {\gamma} (1-{\beta}^2 {\gamma}^2)\right ).
\end{align}

\bibliographystyle{aipnum4-1}
\bibliography{aipsamp} 
\end{document}